\documentclass[journal, 10pt, twocolumn]{IEEEtran}

\usepackage{epsf}
\usepackage{graphics}
\usepackage{times}
\usepackage{graphicx}
\usepackage{amsmath}
\usepackage{amssymb}
\usepackage{epsfig}
\usepackage{subfigure}
\usepackage{array}
\usepackage{float}
\usepackage{algorithm}
\usepackage{algpseudocode}

\begin{document}

\title{Low Complexity Suboptimal ML Detection for OFDM-IM Systems}

\author{Kee-Hoon Kim
\thanks{The author is with the School of Electronic and Electrical Engineering and
IITC, Hankyong National University, Anseong 17579, South Korea (e-mail:
keehk85@gmail.com)}}
\maketitle

\begin{abstract}
Orthogonal frequency division multiplexing with index modulation (OFDM-IM) is a novel multicarrier scheme, which uses the $k$ out of $n$ subcarriers as active subcarriers to transmit data. For detecting the subcarrier activation pattern (SAP) at the receiver, maximum likelihood (ML) detection cannot be used because of its high computational complexity. Instead, the detector selecting the most likely active $k$ subcarriers is used, which is called a $k$ largest values ($k$lv) detector. However, this method degrades the detection performance especially if the ratio of illegal SAPs to SAPs is high. In this letter, the suboptimal ML detector is proposed, which is a slight modification of the $k$lv detector. However, the proposed detector has a similar detection performance compared to the ML detection, which is suitable for flexible implementation of OFDM-IM systems.
\end{abstract}


\section{Introduction}

Orthogonal frequency division multiplexing with index modulation (OFDM-IM) \cite{bacsar2013orthogonal} is a novel multicarrier technique, which extends the concept of spatial modulation (SM) \cite{mesleh2008spatial} into frequency domain. In OFDM-IM, the subcarriers are partitioned into a series of subblocks.
Also, the information bits are conveyed by not only the modulated symbols but also the subcarrier indices unlike the conventional OFDM. That is, the subcarriers have two states, active and inactive, and the indices of the active subcarriers carry information.
The special design of OFDM-IM reduces inter-carrier interference (ICI) and gives better bit error rate (BER) performance in the low to medium data rate region than the conventional OFDM \cite{bacsar2013orthogonal}. Also, it is possible to generate energy efficient signals compared to the conventional OFDM \cite{zhao2012high}.

For detecting the subcarrier activation pattern (SAP) at the receiver, the optimal method is maximum likelihood (ML) detection, where it detects jointly both the indices of the active subcarriers and the modulated symbols carried on. However, naive implementation of the ML detector requires a huge computational complexity.

In \cite{zheng2015low, zhang2017dual}, by using the fact that each symbol can be demodulated independently, the equivalent ML detector is proposed, which only needs to search through all possible realizations of SAP and the $M$ signal space
for each symbol, leading to a reduced computational complexity. In spite of the investigation in \cite{zheng2015low, zhang2017dual}, this ML detector would still become impractical if the number of possible SAPs is large.

To solve this problem, one can practically employ a low-complexity near ML detector which simply picks up $k$ active indices that have $k$ largest values of active likelihood metrics, called a $k$ largest values ($k$lv) detector in this letter.
However, the $k$lv detector may also decide on an illegal SAP that do not belong to the set of the legal SAPs, resulting in degraded detection performance.
The authors in \cite{zheng2015low} mentioned that the probability of this event is very small and thus the performance loss is negligible. However, as the ratio of illegal SAPs to SAPs increases, the degradation of the detection performance of this $k$lv detector cannot be ignored.

In this letter, the suboptimal ML detector for OFDM-IM is proposed. The suboptimal ML detector is a slight modification of the $k$lv detector and thus has likewise low complexity. However, its detection performance is almost the same as the ML detector, as verified thorough the probabilistic analysis and simulation results. By using the proposed suboptimal ML detector, OFDM-IM systems can be implemented with low complexity and suboptimal detection performance.

\subsection{OFDM-IM}

In the OFDM-IM system using $N$ subcarriers, $m$ information bits enter the OFDM-IM transmitter for transmission of one OFDM-IM block. These $m$ bits are divided into $G$ groups, where each contains $p$ bits, i.e., $m=pG$. The $p$ bits in each group are mapped to one subblock of length $n$ in frequency domain, where $n = N/G$.
Unlike the conventional OFDM, this mapping procedure is not only performed by assigning the corresponding modulated symbols, but also by the indices of the subcarriers \cite{bacsar2013orthogonal}.

Specifically, for each subblock, only $k$ out of $n$ subcarriers are activated and the pattern is determined based on the first $p_1$ bits of the $p$ bits in the group. The remaining $p_2=k \log_2M$ bits of the $p$ bits, i.e., $p = p_1 + p_2$, are mapped onto the $M$-ary signal constellation to determine the symbols in the active subcarriers. We set the symbols in the inactive subcarriers to zero. In other words, in the OFDM-IM system, the information is conveyed by both of the $M$-ary modulated symbols and the indices of the active subcarriers \cite{bacsar2013orthogonal}.
Since the number of possible patterns is $\binom{n}{k}$, there has to be $\binom{n}{k}-2^{p_1}$ redundancy or illegal SAPs.
We denote the set of the $\binom{n}{k}$ possible SAPs as $\mathcal{I}$.
Also we denote the set of the $2^{p_1}$ legal SAPs as $\mathcal{I}_l$ and denote the set of the $\binom{n}{k}-2^{p_1}$ illegal SAPs as $\mathcal{I}_i$. Clearly, $\mathcal{I} = \mathcal{I}_l \cup \mathcal{I}_i$.

Denote the set of the indices of the $k$ active subcarriers in the transmitted $g$-th OFDM-IM subblock, $g=1,2,\cdots,G$, as
\begin{equation}
I_{g} = \{i_{g,1},i_{g,2},\cdots,i_{g,k}\}
\end{equation}
with $i_{g,m} \in \{1,2,\cdots,n\}$ for $m = 1,2,\cdots,k$. Clearly, $I_{g} \in \mathcal{I}_l$.
Correspondingly, the set of $k$ modulated symbols is denoted by
\begin{equation}
S_{g} =\{S_{g,1},S_{g,2},\cdots,S_{g,k}\},
\end{equation}
where $S_{g,m} \in \mathcal{S}$ and $\mathcal{S}$ is the used signal constellation.
Then the $g$-th OFDM-IM subblock can be constructed as
\begin{equation}
\mathbf{X}_{g} = [X_{g,1}~X_{g,2}~\cdots~X_{g,n}]^T,
\end{equation}
where the $i$-th OFDM-IM symbol $X_{g,i} \in \mathcal{S}$ only if $i \in I_{g}$ and otherwise $X_{g,i} = 0$.

The OFDM-IM transmitter creates $\mathbf{X}_{g}$ for all $g$.
Then the $G$ subblocks are concatenated to generate the $N\times 1$ OFDM-IM symbol sequence. For achieving frequency diversity gain as much as possible, concatenation in an interleaved manner is employed.
After these point, the same procedure as the conventional OFDM is applied. The symbol sequence in frequency domain is processed by the inverse discrete Fourier transform (IDFT) to generate the OFDM-IM signal in time domain. Then cyclic prefix (CP) is appended followed by parallel-to-serial (P/S) and digital-to-analog (D/A) conversion.

\subsection{Detection for OFDM-IM}

Let us consider the detection of the $g$-th subblock. We omit the subblock index $g$ for simplicity. By considering a joint detection for the indices of the active subcarriers and the modulated symbols carried on, the ML detector for OFDM-IM is given by
\begin{align}\label{eq:MLd}
\{\hat{I}_\mathrm{ML}, \hat{S}\}
&= \arg\min_{\tilde{I} \in \mathcal{I}_l,\tilde{S}}  \sum_{i=1}^{n}|Y_{i}-H_{i}X_{i}|^2\nonumber\\
&= \arg\min_{\tilde{I} \in \mathcal{I}_l ,\tilde{S}}  \sum_{i=1}^{n} |H_{i}|^2 |R_i-X_{i}|^2,
\end{align}
where $Y_{i} = H_iX_i + Z_i$ is the $i$-th received OFDM-IM symbol, $H_{i}$ is the $i$-th channel frequency response (CFR), $Z_i$ is the Gaussian noise with $\mathcal{CN}(0,2\sigma^2)$, and $R_i = H^{-1}_{i}Y_{i}$ for $i=1,\cdots,n$.

It is remarkable that the symbol detection can be independently performed for each subcarrier \cite{zheng2015low, zhang2017dual}. Then, the symbol detection is separately performed as
\begin{equation}
\hat{s}_{i} = \arg\min_{s \in \mathcal{S}}|R_i-s|^2
\end{equation}
for $i = 1,\cdots, n$. Then, (\ref{eq:MLd}) becomes
\begin{equation}\label{eq:Iselect}
\hat{I}_\mathrm{ML} = \arg\min_{\tilde{I} \in \mathcal{I}_l} \left\{ \sum_{i\in \tilde{I} } |H_{i}|^2 |R_i-\hat{s}_{i}|^2 + \sum_{j\notin \tilde{I} } |H_{j}|^2 |R_j|^2\right\}.
\end{equation}

Since $\sum_{i=1}^{n} |H_{i}|^2 |R_i|^2$ is not related to the realizations of $\tilde{I}$, we subtract it from (\ref{eq:Iselect}). Then we have
\begin{align}\label{eq:MLdetect}
\hat{I}_\mathrm{ML}
&= \arg\min_{\tilde{I}\in \mathcal{I}_l}  \sum_{i\in \tilde{I}} |H_{i}|^2 (|R_{i}-\hat{s}_{i}|^2 - |R_{i}|^2)\nonumber\\
&= \arg\max_{\tilde{I}\in \mathcal{I}_l}  \sum_{i\in \tilde{I}} |H_{i}|^2 (|R_{i}|^2 - |R_{i}-\hat{s}_{i}|^2)\nonumber\\
&= \arg\max_{\tilde{I}\in \mathcal{I}_l}  \sum_{i\in \tilde{I}} A_i,
\end{align}
where
\begin{align}\label{eq:almA}
A_i &= |H_{i}|^2 (|R_{i}|^2 - |R_{i}-\hat{s}_{i}|^2)\nonumber\\
&= |H_{i}|^2 (2\mathrm{Re}\{R_i^*\hat{s}_{i}\}  -|\hat{s}_{i}|^2)
\end{align}
is an active likelihood metric for the $i$-th subcarrier.

Since the ML detector calculates $2^{p_1}$ combinations of $A_i$ in (\ref{eq:MLdetect}), the ML detector would become impractical for a larger $p_1$ as $2^{p_1}$ grows exponentially with it. Therefore, the $k$lv detector that chooses the indices with the $k$ largest values of $A_i$ may be preferred in practical systems.
That is, the $k$lv detector is
\begin{equation}\label{eq:klvdetect}
\hat{I}_{k\mathrm{lv}} = \arg\max_{\tilde{I}\in \mathcal{I}}  \sum_{i\in \tilde{I}} A_i.
\end{equation}

This $k$lv detector may also decide on illegal SAPs that do not belong to $\mathcal{I}_l$, resulting in degraded detection performance. Although the probability of this error event is small unless the ratio of illegal SAPs to SAPs is large \cite{zheng2015low}, this constraint prevents the flexible implementation of OFDM-IM systems with various parameters $n$ and $k$.

\section{The Proposed Suboptimal ML Detection}
\subsection{Active Likelihood Metric $A_i$}\label{sec:ALM}
If we employ quadrature phase shift keying (QPSK) for modulating symbols, $A_i$ in (\ref{eq:almA}) becomes
\begin{equation}
A_i = 2|H_i|^2(|\mathrm{Re}\{R_i\}|+|\mathrm{Im}\{R_i\}|-1).
\end{equation}
For a given $H_i$, $A_i$ is a Gaussian distribution with $\mathcal{N}(|H_i|^2,2|H_i|^4\sigma^2)$ if the $i$-th subcarrier is active. Otherwise, $A_i$ becomes a distribution of $\mathcal{N}(-|H_i|^2,2|H_i|^4\sigma^2)$.
Assume that the $i$-th subcarrier is active and the $j$-th subcarrier
is inactive. Since the means of $A_i$ and $A_j$ are opposite to
each other, confused detection of the $i$-th and $j$-th
subcarriers occurs when $A_i$ and $A_j$ are close to zero. It means that bad channel qualities $H_i$ and $H_j$ at the same
time are necessary for confused detection of the $i$-th and $j$-th
subcarriers. This phenomenon can also be mentioned in \cite{bacsar2013orthogonal}, where it is shown that the index demodulation error event has a diversity order of two.

For future use, we denote the indices of $A_i$ as $\hat{i}_{1},\cdots, \hat{i}_{n}$ when $A_i$ are sorted in descending order. That is,
\begin{equation}\label{eq:met}
A_{\hat{i}_{1}} > A_{\hat{i}_{2}} > \cdots > A_{\hat{i}_{n}}.
\end{equation}
Then, the set constructed by the indices of the $k$ largest values of $A_i$ becomes the best SAP of the $k$lv detector in (\ref{eq:klvdetect}) as
\begin{equation}
\hat{I}_{k\mathrm{lv}} =\hat{I}_1 = \{ \hat{i}_{1}, \hat{i}_{2}, \cdots, \hat{i}_{k}\}.
\end{equation}
We may also denote $\hat{I}_v$'s for $v=2,\cdots,\binom{n}{k}$, which means the $v$-th best SAP based on the metrics in (\ref{eq:met}).
Clearly, the second best SAP $\hat{I}_2$ is
\begin{equation}\label{eq:secSAP}
\hat{I}_2 = \{ \hat{i}_{1}, \hat{i}_{2}, \cdots, \hat{i}_{k-1}, \hat{i}_{k+1}\}.
\end{equation}
Note that the other $v$-th best SAPs ($v\geq 3$) are not fixed and can be varied according to the specific values of $A_i$'s.
For example, the third best SAP $\hat{I}_3$ is either $\{ \hat{i}_{1}, \hat{i}_{2}, \cdots, \hat{i}_{k-1}, \hat{i}_{k+2}\}$ or $\{ \hat{i}_{1}, \hat{i}_{2}, \cdots, \hat{i}_{k-2}, \hat{i}_{k}, \hat{i}_{k+1}\}$ according to the values of $A_i$.

\subsection{Correct Detection Probabilities of $\hat{I}_{k\mathrm{lv}}$ and $\hat{I}_{\mathrm{ML}}$}

\begin{figure}[htbp]
\centering
\includegraphics[width=.7\linewidth]{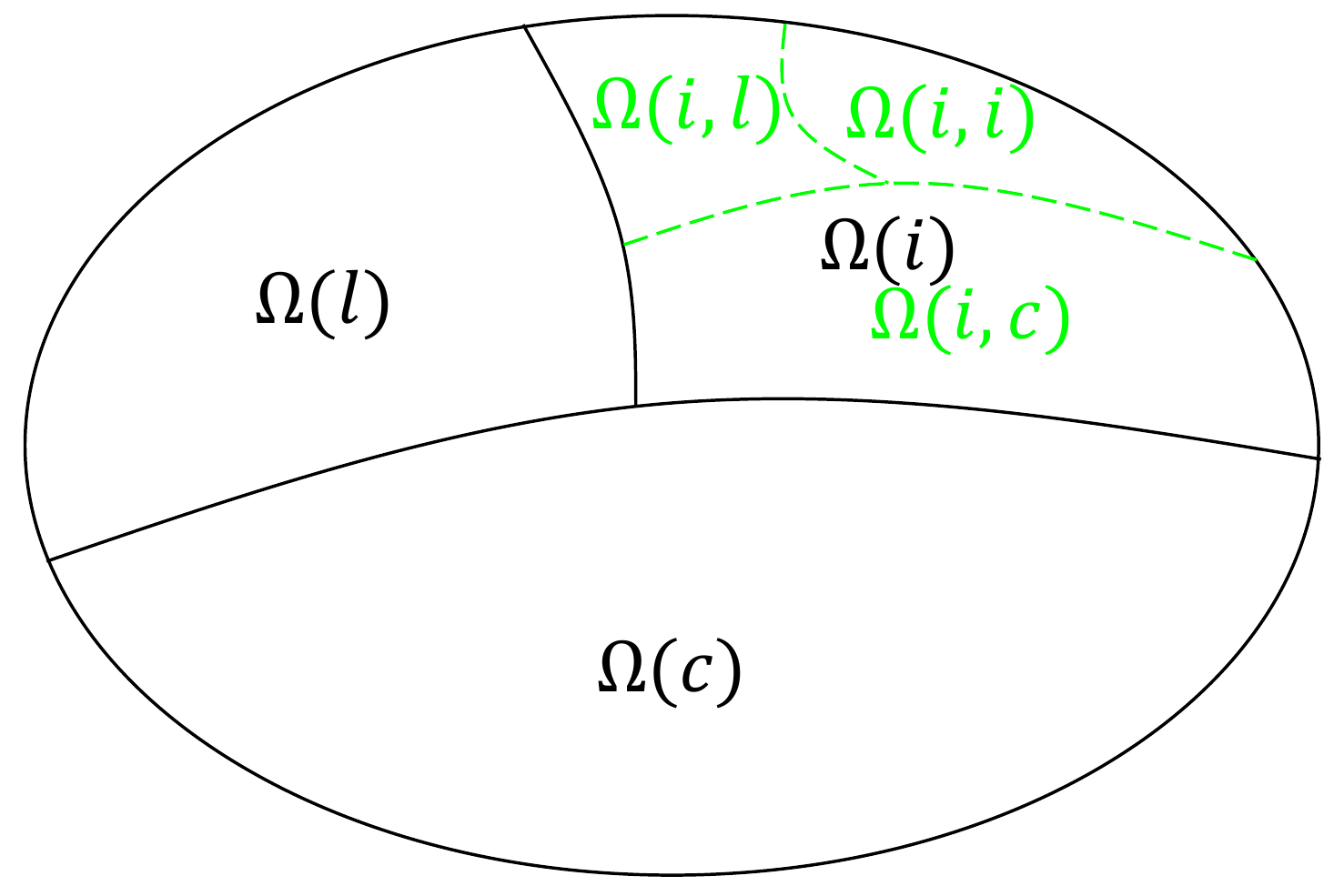}
\caption{A sample space of a received OFDM-IM subblock.}
\label{fig:ss}
\end{figure}

Consider a sample space in probability theory of the received OFDM-IM subblock, which denotes the set of all possible realizations.
The sample space can be separated into three sets according to which the best SAP $\hat{I}_1$ is, as in Fig. \ref{fig:ss}.

Specifically, the sets are separated by the following criteria:
\begin{itemize}
  \item $\Omega(c)$: The best SAP is correct. ($\hat{I}_1 = I$)
  \item $\Omega(l)$: The best SAP is incorrect and legal. ($\hat{I}_1 \neq I$ and $\hat{I}_1 \in \mathcal{I}_l$)
  \item $\Omega(i)$: The best SAP is incorrect and illegal. ($\hat{I}_1 \in \mathcal{I}_i$)
 \end{itemize}
Moreover, according to the second best SAP $\hat{I}_2$, $\Omega(i)$ can be separated into three subsets as
\begin{itemize}
  \item $\Omega(i,c)$: $\hat{I}_1 \in \mathcal{I}_i$ and the second best SAP $\hat{I}_2$ is correct.
  \item $\Omega(i,l)$: $\hat{I}_1 \in \mathcal{I}_i$ and the second best SAP $\hat{I}_2$ is incorrect and legal.
  \item $\Omega(i,i)$: $\hat{I}_1 \in \mathcal{I}_i$ and the second best SAP $\hat{I}_2$ is incorrect and illegal.
\end{itemize}
Likewise, $\Omega(i,i)$ can be further separated into three subsets $\Omega(i,i,c), \Omega(i,i,l)$, and $\Omega(i,i,i)$ according to the third best SAP $\hat{I}_3$.
For example, $\Omega(i,i,c)$ means $\hat{I}_1 \in \mathcal{I}_i$, $\hat{I}_2 \in \mathcal{I}_i$, and $\hat{I}_3 = I$.
In the same manner, this separation can be performed until we have  $\Omega(\underbrace{i,i,\cdots,i}_{\binom{n}{k}-2^{p_1}},c)$.

Clearly, the correct detection probability of the $k$lv detector is
\begin{equation}\label{eq:klv}
P_{k\mathrm{lv}} = P(\Omega(c)).
\end{equation}
The ML detector in (\ref{eq:MLdetect}) finds the SAP having the largest sum of $A_i$ in the set of legal SPAs $\mathcal{I}_l$ as in (\ref{eq:MLdetect}). Therefore, if we use the ML detector, then not only the case in $\Omega(c)$ but also the cases in $\Omega(i,c) + \cdots + \Omega(\underbrace{i,i,\cdots,i}_{\binom{n}{k}-2^{p_1}},c)$
can be correctly detected by the ML detector.
That is, the correct detection probability of the ML detector is
\begin{equation}\label{eq:ML}
P_{\mathrm{ML}} = P(\Omega(c)) + P(\Omega(i,c)) + \cdots + P(\Omega(\underbrace{i,i,\cdots,i}_{\binom{n}{k}-2^{p_1}},c)).
\end{equation}

Therefore, the ML detection is superior to the $k$lv detector.
Also, from (\ref{eq:klv}) and (\ref{eq:ML}), the probability gap becomes
\begin{align}\label{eq:gapMLklv}
P_{\mathrm{ML}} - P_{k\mathrm{lv}}
&= P(\Omega(i,c)) + \cdots + P(\Omega(\underbrace{i,i,\cdots,i}_{\binom{n}{k}-2^{p_1}},c))\nonumber\\
&\leq P(\Omega(i))\nonumber\\
&=\frac{\binom{n}{k} - 2^{p_1}}{\binom{n}{k}-1}\cdot(1-P(\Omega(c)))\nonumber\\
&=r\cdot(1-P(\Omega(c))),
\end{align}
where $r$ is the ratio of the illegal SAPs to the all incorrect SAPs as
\begin{equation}
r = \frac{\binom{n}{k} - 2^{p_1}}{\binom{n}{k}-1}.
\end{equation}

Without loss of generality, we consider the transmitted SAP $I=\{1,2,\cdots,k\}$.
Then,
\begin{equation}
P(\Omega(c)) = P(\min(A_1,\cdots,A_{k})>\max(A_{k+1},\cdots,A_n)),
\end{equation}
where the probability $P(\Omega(c))$ is regardless of $r$. Therefore, the gap $P_{\mathrm{ML}} - P_{k\mathrm{lv}}$ in (\ref{eq:gapMLklv}) becomes larger as $r$ increases.

\subsection{The Proposed Suboptimal ML Detector}

We focus on the fact that in (\ref{eq:ML}) the first and second terms are dominant and these terms can be obtained when we also test the second best SAP in addition to the first best SAP.
Fortunately, the second best SAP is fixed as in (\ref{eq:secSAP}).
Using these, we propose the suboptimal ML detector in Algorithm \ref{al:subML}.
\begin{algorithm}
\caption{Suboptimal ML Detection}\label{al:subML}
\begin{algorithmic}[1]
   \State $\hat{I}_1 = \{\hat{i}_1, \hat{i}_2, \cdots, \hat{i}_k\} $
   \State $\hat{I}_2 = \{\hat{i}_1, \hat{i}_2, \cdots, \hat{i}_{k-1}, \hat{i}_{k+1}\} $\Comment{Newly added}
   \If{$\hat{I}_1 \in \mathcal{I}_l$}
      \State $\hat{I}_{\mathrm{subML}} \gets \hat{I}_1$
   \ElsIf{$\hat{I}_2 \in \mathcal{I}_l$}\Comment{Newly added}
     \State $\hat{I}_{\mathrm{subML}} \gets \hat{I}_2$\Comment{Newly added}
   \EndIf
   \State \textbf{return} $\hat{I}_{\mathrm{subML}}$
\end{algorithmic}
\end{algorithm}

Note that the proposed suboptimal ML detector is a slight modification of the $k$lv detector in (\ref{eq:klvdetect}) and the parts newly added are marked in Algorithm \ref{al:subML}. 
After calculating and sorting the values of $A_i$ for $i=1,\cdots,n$, the rest procedure of the $k$lv detector is investigation $\hat{I}_1 \in \mathcal{I}_l$ as in the third line in Algorithm \ref{al:subML}. The computational complexity of this investigation procedure is negligible because the SAP $\hat{I}_1$ can be seen as a binary representation. 
Clearly, the added parts of the proposed ML detector induce no additional complexity burden because $\hat{I}_2$ is fixed as in (\ref{eq:secSAP}) and the computational complexity of the investigation procedure of $\hat{I}_2 \in \mathcal{I}_l$ is negligible as $\hat{I}_1 \in \mathcal{I}_l$.

If we use the proposed suboptimal ML detector, then the received OFDM-IM subblock in
$\Omega(c)$ and $\Omega(i,c)$ in Fig. \ref{fig:ss}
can be correctly detected. Then its correct detection probability is
\begin{equation}\label{eq:subML}
P_{\mathrm{subML}} = P(\Omega(c)) + P(\Omega(i,c)).
\end{equation}
The difference between (\ref{eq:subML}) and (\ref{eq:ML}) is
\begin{align}\label{eq:diff}
P_{\mathrm{ML}} - P_{\mathrm{subML}} &= P(\Omega(i,i,c) + \cdots + \Omega(i,i,\cdots,i,c))\nonumber\\
&\leq P(\Omega(i,i))\nonumber\\
&= \frac{\binom{n}{k} - 2^{p_1}-1}{\binom{n}{k}-1}\cdot (P(\Omega(i))-P(\Omega(i,c))).
\end{align}

Now we consider $P(\Omega(i))$ and $P(\Omega(i,c))$ in (\ref{eq:diff}).
Without loss of generality, we assume that the transmitted SAP is $I=\{1,2,\cdots,k\}$.
First, $P(\Omega(i))$ becomes
\begin{align}\label{eq:Omegai}
&P(\Omega(i))\nonumber\\
&= P(\hat{I}_1\in \mathcal{I}_i)\nonumber\\
&= P(\hat{I}_1\in \mathcal{I}_i \cap |\hat{I}_1-I| = 2) + P(\hat{I}_1\in \mathcal{I}_i \cap |\hat{I}_1-I| = 4) + \cdots\nonumber\\
&\simeq P(\hat{I}_1\in \mathcal{I}_i \cap |\hat{I}_1-I| = 2)\nonumber\\
&= r\cdot  P(|\hat{I}_1-I| = 2)\nonumber\\
&= r \cdot k(n-k)\cdot P(\hat{I}_1 = \{1,\cdots,k-1,k+1\}) \nonumber\\
&= r \cdot k(n-k)\nonumber\\
&\cdot P(\min(A_1,\cdots,A_{k-1},A_{k+1})>\max(A_k,A_{k+2},\cdots,A_n)),
\end{align}
where the similarity in the third line is reasonable because the event $|\hat{I}_1-I|=2$ frequently occurs compared to the other events and $k(n-k)$ in the fifth line comes from the number of $\hat{I}_1$ satisfying $|\hat{I}_1-I| = 2$.

In the similar way, we also have
\begin{align}\label{eq:Omegaic}
&P(\Omega(i,c))\nonumber\\
& = P(\hat{I}_1 \in \mathcal{I}_i \cap  \hat{I}_2 = I)\nonumber\\
& \simeq P(\hat{I}_1 \in \mathcal{I}_i \cap  \hat{I}_2 = I \cap |\hat{I}_1-I| = 2)\nonumber\\
& = r\cdot P(\hat{I}_2 = I \cap |\hat{I}_1-I| = 2)\nonumber\\
& = r\cdot k(n-k)\cdot P(\hat{I}_1 = \{1,\cdots,{k-1},{k+1}\} \cap \hat{I}_2 = \{1,\cdots,k\})\nonumber\\
&= r\cdot k(n-k)\nonumber\\
&\cdot P(\min(A_1,\cdots,A_{k-1})>A_{k+1}>A_k>\max(A_{k+2},\cdots,A_n)).
\end{align}

From (\ref{eq:Omegai}) and (\ref{eq:Omegaic}), $P(\Omega(i))-P(\Omega(i,c))$ becomes
\begin{align}\label{eq:UVA}
& P(\Omega(i))-P(\Omega(i,c))\nonumber\\
&\simeq  r\cdot  k(n-k)\cdot (P(\min(U,A_{k+1})>\max(A_k, V))\nonumber\\
&~~~~~~~~~~~~~~~~~~~~- P(U>A_{k+1}>A_k>V))\nonumber\\
&=  r\cdot  k(n-k)\cdot(P(A_{k+1}>U>A_k>V)\nonumber\\
&~~~~~~~~~~~~~~~~~~~~ + P(U>A_{k+1}>V>A_k) \nonumber\\
&~~~~~~~~~~~~~~~~~~~~ + P(A_{k+1}>U>V>A_k)),
\end{align}
where
\begin{align}
U &= \min(A_1,\cdots,A_{k-1})\\
V &= \max(A_{k+2},\cdots,A_n).
\end{align}

Let us consider the three probabilities in (\ref{eq:UVA}).
Note that the bad channel qualities are necessary condition for
confusion of active subcarriers, as explained in subSection \ref{sec:ALM}.
Then the event $A_{k+1}>U>A_k>V$ in (\ref{eq:UVA}) occurs rarely because this event requires that the $k$-th, $k+1$-th, and $z$-th ($1\leq z \leq k-1$) CFRs are bad at the same time. That is, this event has a frequency diversity order of three.
Likewise, the other two events $U>A_{k+1}>V>A_k$ and $A_{k+1}>U>V>A_k$ require three and four bad CFRs, respectively.
Therefore, $P(\Omega(i)-\Omega(i,c))$ in (\ref{eq:UVA}) is small and thus, from (\ref{eq:diff}), we expect that the detection performance gap between the ML detector and the proposed suboptimal ML detector is also small especially in high signal-to-noise ratio (SNR) region.

\section{Simulation Results}
To verify the performance of the proposed suboptimal ML detector, we simulate two OFDM-IM systems with two different illegal SAPs ratios $r$. For modulating the symbols in the active subcarriers, QPSK is commonly used because OFDM-IM gives better BER performance in the low to medium data rate region than the conventional OFDM \cite{bacsar2013orthogonal}. Also, we consider a Rayleigh fading channel with length eight having the exponential power-delay profile. Since an interleaved concatenation is employed, in frequency domain, the elements within an OFDM-IM subblock experience nearly independent CFRs.

\begin{figure}[htbp]
\centering
\includegraphics[width=.9\linewidth]{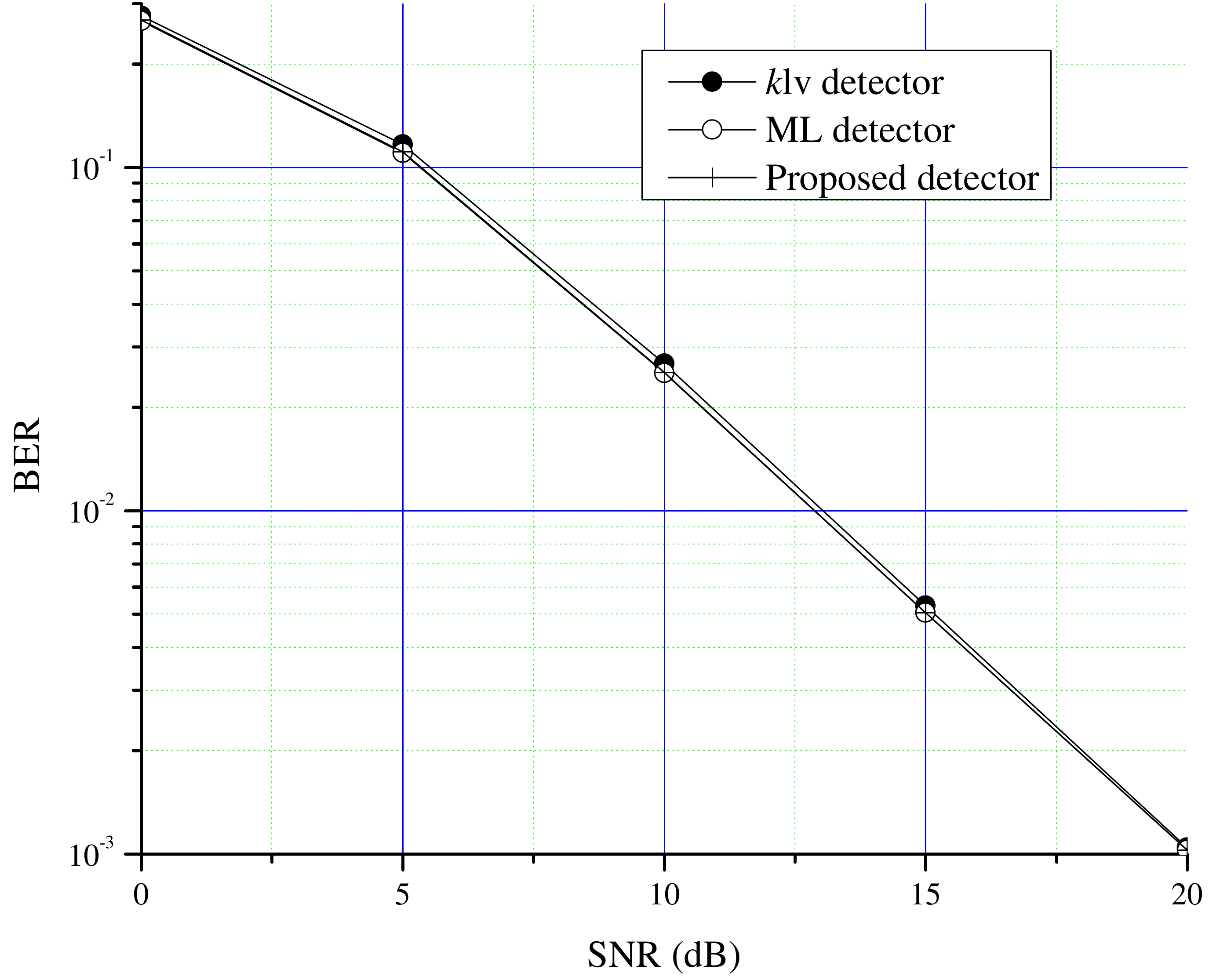}
\caption{BER performance of the three detectors, where we use $N=128$, $n=8$, and $k=4$.}
\label{fig:n8k4}
\end{figure}

\begin{figure}[htbp]
\centering
\includegraphics[width=.9\linewidth]{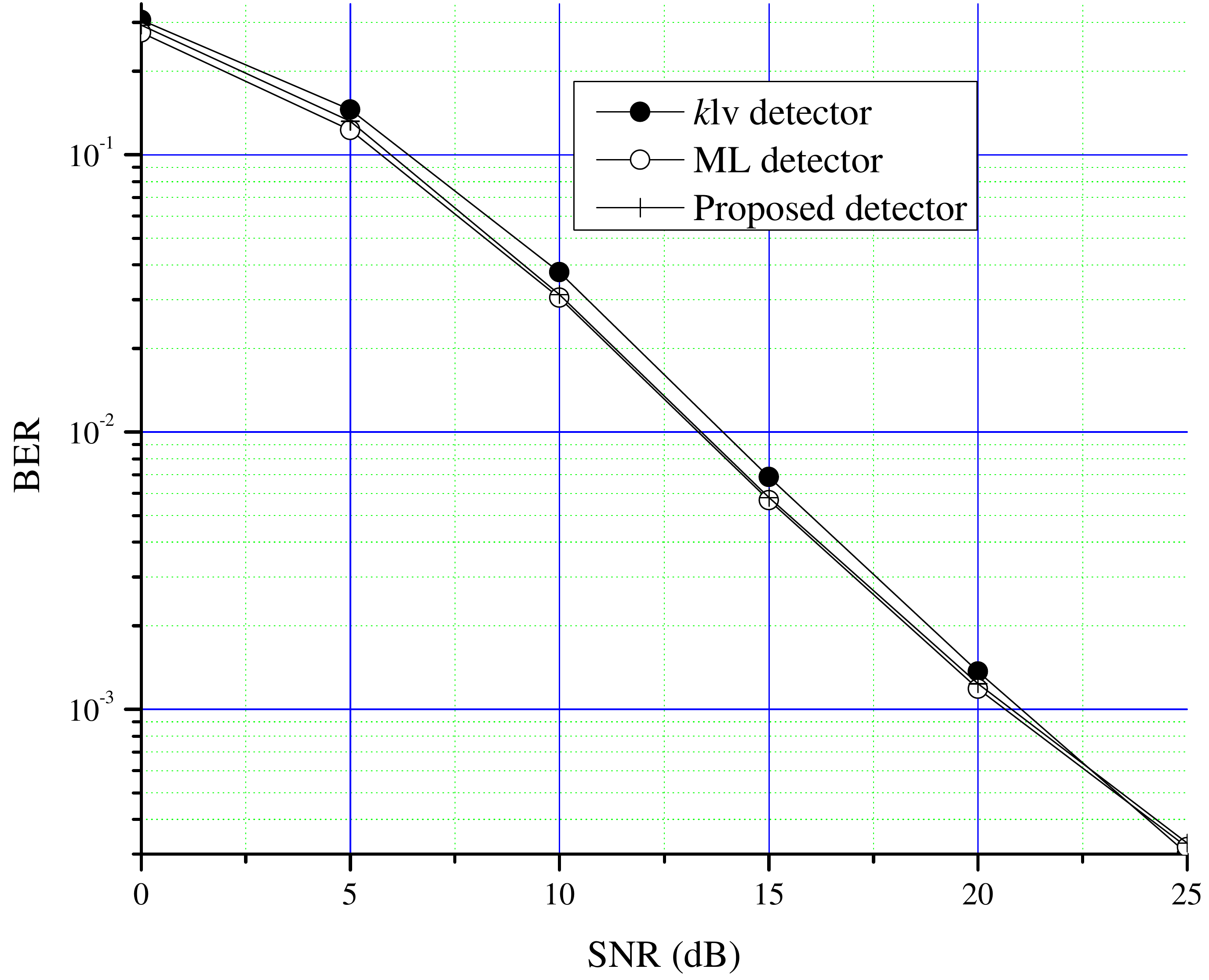}
\caption{BER performance of the three detectors, where we use $N=100$, $n=10$, and $k=5$.}
\label{fig:n10k5}
\end{figure}

Fig. \ref{fig:n8k4} shows the BER performance of the three detectors, where we use $N=128$, $n=8$, and $k=4$.
In this case, the illegal SAPs ratio is only $r= 0.086$ and thus there is only a small gap between the ML detector and the $k$lv detector, described in (\ref{eq:gapMLklv}). Also, the proposed suboptimal ML detector shows a similar BER performance compared to the ML detector, explained in (\ref{eq:UVA}).
The error events within different subblocks are identical and it is sufficient to investigate the error events within a single subblock to determine the overall system performance. Therefore, it is enough to verify the OFDM-IM systems with small size of $N$.

Fig. \ref{fig:n10k5} shows the BER performance of the three detectors, where we use $N=100$, $n=10$, and $k=5$.
In this case, the redundancy SAPs ratio is $r= 0.49$ and thus there is a visible gap between the ML detector and the $k$lv detector. Then, the proposed suboptimal ML detector shows almost the same BER performance compared to the ML detector. As explained in (\ref{eq:UVA}), the performance gap between the ML detector and the proposed detector becomes smaller as SNR increases.

\section{Conclusion}
In this letter, the suboptimal ML detection for OFDM-IM systems is proposed, where the second best SAP is subsequently tested after the test of the first best SAP.
This simple modification can significantly enhance the detection performance because it is enough for boosting the detection performance to test the first and second best SAPs only, which is analyzed.
By using the proposed suboptimal ML detector with low complexity, we obtain almost the same detection performance compared to the ML detector.
This leads to the flexible and unconstrained implementation of OFDM-IM systems.

\bibliographystyle{IEEEtran}
\bibliography{biblio,IEEEfull}

\end{document}